\documentclass[aps,prd,twocolumn,notitlepage,superscriptaddress,nofootinbib,floatfix]{revtex4-2}
\usepackage[utf8]{inputenc} %Support for additional characters in bibliography
\usepackage{amsfonts,amssymb, mathrsfs, amsmath, esint,bm,enumitem}
\usepackage{subcaption}
\allowdisplaybreaks
\usepackage{ragged2e}
\DeclareCaptionJustification{justified}{\justifying}
\makeatletter    
\renewcommand\onecolumngrid{% <<<<<<
\do@columngrid{one}{\@ne}%
\def\set@footnotewidth{\onecolumngrid}% <<<<<<<<<<<<<<<<
\def\footnoterule{\kern-6pt\hrule width 1.5in\kern6pt}%
}
\renewcommand\twocolumngrid{% <<<<<<
        \def\footnoterule{% restore rule
        \dimen@\skip\footins\divide\dimen@\thr@@
        \kern-\dimen@\hrule width.5in\kern\dimen@}
        \do@columngrid{mlt}{\tw@}
}%
\makeatother 

\usepackage{diagbox}
\usepackage{latexsym}
\usepackage{graphicx}
\usepackage[dvipsnames]{xcolor}
\usepackage{booktabs,multirow}
\usepackage{titlesec} % to change style of \paragraph{} header
\usepackage{physics}
\usepackage[normalem]{ulem}
\graphicspath{{./Figures/}}

\usepackage{hyperref}
\hypersetup{colorlinks, citecolor=ForestGreen, linkcolor=BrickRed, urlcolor=RedViolet}

\renewcommand{\d}{\mathrm{d}}

\newcommand{\QCD}{\text{\tiny QCD}}
\newcommand{\MP}{M_\textsc{p}}

\newcommand{\Hz}{\,\text{Hz}}
\newcommand{\nHz}{\,\text{nHz}}
\newcommand{\GeV}{\,\text{GeV}}
\newcommand{\MeV}{\,\text{MeV}}

\newcommand{\gs}{g_\star}

\newcommand{\nt}{n_T}
\renewcommand{\S}{\mathcal S}
\renewcommand{\cp}{\textsc{cp}}
\newcommand{\yr}{\text{yr}}
\newcommand{\ct}{\textsc{ct}}
\newcommand{\cgw}{\textsc{cgw}}
\newcommand{\gw}{\textsc{gw}}
\newcommand{\ds}{s}

\newcommand{\OGW}{\Omega_\gw}
\newcommand{\fp}{f_\star}
\newcommand{\fpk}{\fp}
\newcommand{\Neff}{N_\text{eff}}
\newcommand{\FS}{\textsc{fs}}
\newcommand{\fFS}{f_\FS}
\def\parfrac#1#2{{\left(\frac{#1}{#2}\right)}}

\newcommand{\FRcol}[1]{\textcolor{Cyan}{#1}}
\renewcommand{\FR}[1]{\FRcol{\textsf{(FR: #1)}}}

\begin{document}
% TITLE< AUTHORS, AFFILIATIONS AND ABSTRACTS FOR PRL
\title{
Footprints of the QCD Crossover on Cosmological Gravitational Waves \\
at Pulsar Timing Arrays
}

\author{Gabriele Franciolini} 
\email{gabriele.franciolini@uniroma1.it}
\affiliation{Dipartimento di Fisica, Sapienza Università di Roma and INFN, Sezione di Roma,\\
Piazzale Aldo Moro 5, 00185, Rome, Italy}
\author{Davide~Racco} 
\email{dracco@stanford.edu}
\affiliation{Stanford Institute for Theoretical Physics, Stanford University, 382 Via Pueblo Mall, Stanford, CA 94305, U.S.A.}
\author{Fabrizio Rompineve} 
\email{fabrizio.rompineve@cern.ch}
\affiliation{CERN, Theoretical Physics Department,
Esplanade des Particules 1, Geneva 1211, Switzerland} 
\affiliation{Departament de F\'isica, Universitat Aut\`onoma de Barcelona, 08193 Bellaterra, Barcelona, Spain
\looseness=-1}
\affiliation{Institut de F\'isica d’Altes Energies (IFAE) and The Barcelona Institute of Science and Technology (BIST), Campus UAB, 08193 Bellaterra (Barcelona), Spain}%\looseness=-1}

\begin{abstract}
\noindent
Pulsar Timing Arrays (PTAs) have reported evidence for a stochastic gravitational wave (GW) background at nHz frequencies, possibly originating in the early Universe. We show that the spectral shape of the low-frequency (causality) tail of GW signals sourced at temperatures around $T\gtrsim 1$ GeV is distinctively affected by confinement of strong interactions (QCD), due to the corresponding sharp decrease in the number of relativistic species. 
Bayesian analyses in the NANOGrav 15 years and the previous International PTA datasets reveal a significant improvement in the fit with respect to cubic power-law spectra, previously employed for the causality tail. This suggests that the inclusion of Standard Model effects on GWs can have a potentially decisive impact on model selection. 
\end{abstract}

\hfill CERN-TH-2023-080

\maketitle
\vspace{1em}\noindent

%%%%%%%%%%%%%%%%%%%%%%%%%%%%%%%%%%%%%%%%%%%%%%%%%%%%%%%%%%%%%%%%%%%%%%%%
\section{Introduction}
\label{sec:introduction}
%%%%%%%%%%%%%%%%%%%%%%%%%%%%%%%%%%%%%%%%%%%%%%%%%%%%%%%%%%%%%%%%%%%%%%%

A stochastic background of Gravitational Waves (SGWB) may be the only direct probe into the early stages of cosmological evolution, where it can be produced by physics beyond the Standard Model (SM). 
The recently-reported evidence for a nHz SGWB in the NANOGrav 15 years~\cite{NG15-pulsars,NG15-SGWB} (NG15), European PTA Data Release 2 (EPTA-DR2)~\cite{EPTA2-pulsars,EPTA2-SGWB}, Parkes~\cite{PPTA3-pulsars,PPTA3-SGWB} and Chinese~\cite{CPTA-SGWB} PTAs datasets, 
whose uncorrelated component was already detected with previous data~\cite{NANOGrav:2020bcs, Goncharov:2021oub, Chen:2021rqp, Antoniadis:2022pcn},
has attracted ample attention in the astrophysics, cosmology and particle physics communities. 

One of the most urgent endeavors is determining whether the signal is of cosmological or astrophysical origin, in the latter case sourced by supermassive black hole binaries (SMBHBs), see e.g.~\cite{Burke-Spolaor:2018bvk, NANOGrav:2020spf} for an overview. 
This is difficult for multiple reasons. 
First, while the NG15 analysis suggests that the astrophysical model may face challenges~\cite{NG15-SMBHB}, the current understanding of SMBHBs is not sufficiently precise to draw conclusions \cite{Middleton:2020asl,Antoniadis:2022pcn, EPTA2-SMBHB-NP}; Second, the spectrum of a cosmological SGWB generically depends on the microphysical nature of the source and often requires case-by-case numerical simulations.  

In this Letter, we show that a distinctive signature is nonetheless imprinted model-independently by the early-Universe dynamics of the SM, in the GW spectra of a broad class of early Universe sources. 
These are often referred to as causality-limited, i.e.~radiating GWs on time scales comparable to the corresponding Hubble time. 
Examples are: first order phase transitions (PTs)~\cite{Witten:1984rs}, annihilation of cosmic domain walls~\cite{Vilenkin:1981zs}, collapse of large density perturbations~\cite{Matarrese:1997ay} (see also~\cite{Bian:2020urb, NANOGrav:2021flc, Xue:2021gyq, Ferreira:2022zzo, Dandoy:2023jot, Bringmann:2023opz, NG15-NP, EPTA2-SMBHB-NP, Chen:2019xse,  Nakai:2020oit, Vaskonen:2020lbd, DeLuca:2020agl, Neronov:2020qrl, Zhao:2022kvz, RoperPol:2022iel}), and arise in several well-motivated beyond the SM (BSM) scenarios (see instead~\cite{Ellis:2020ena, Blasi:2020mfx, Vagnozzi:2020gtf, Blanco-Pillado:2021ygr, NG15-NP, EPTA2-SMBHB-NP} for PTA-related analyses of other types of possible cosmological SGWBs).

Our starting point is a fortuitous coincidence of scales: the nHz frequencies probed by PTAs coincide with those of GWs that re-enter the Hubble horizon at the epoch of the QCD crossover phase transition, i.e. at temperatures ~$T\sim 100\MeV$ (see e.g.~\cite{Aoki:2006we}). While the crossover is not expected to source GWs, the rapid drop of relativistic degrees of freedom in the thermal bath significantly changes the equation of state (EoS) of the Universe, that is precisely determined by means of lattice QCD~\cite{Borsanyi:2016ksw}. 

A causality-limited GW source, active before the QCD crossover, produces a model-independent low-frequency GW signal, which we refer to as Causality Tail (CT), that is affected by the SM-induced change in the EoS (as similarly pointed out for other GW signals in~\cite{Schwarz:1997gv, Watanabe:2006qe, Schettler:2010dp}), due to the different evolution of GW modes whose wavelength is larger than the Hubble radius when the source is active, see also~\cite{Hook:2020phx, Brzeminski:2022haa,Loverde:2022wih}).
%The CT spectrum is altered bythe redshift of the SM radiation bath (previously pointed out for other GW signals \cite{Watanabe:2006qe, Schettler:2010dp}) and the different evolution of GWs \cite{Hook:2020phx, Brzeminski:2022haa,Loverde:2022wih}.

This Letter derives the spectral shape of CT signals at nHz frequencies, that can be readily used by PTA collaborations, GW and BSM communities for model comparison (improving upon simple power-law CTs, currently adopted by NG15~\cite{NG15-NP} and EPTA-DR2~\cite{EPTA2-SMBHB-NP}, see also~\cite{NANOGrav:2021flc, Xue:2021gyq}). 
Importantly, we show that our novel inclusion of QCD-induced features significantly impacts the interpretation of current PTA data, by performing a Bayesian search for a CT signal in the International PTA data release 2 (IPTA-DR2)~\cite{Perera:2019sca, Antoniadis:2022pcn} and NG15 \cite{NG15-SGWB}.

%%%%%%%%%%%%%%%%%%%%%%%%%%%%%%%%%%%%%%%%%%%%%%%%%%%%%%%%%%%%%%%%%%%%%%%%
\section{Standard Model features in the causality tail of primordial GW backgrounds}
\label{sec:causality tail}
%%%%%%%%%%%%%%%%%%%%%%%%%%%%%%%%%%%%%%%%%%%%%%%%%%%%%%%%%%%%%%%%%%%%%%%

For cosmological SGWBs, a powerful property is ensured in a broad class of primordial sources where GWs are generated locally, independently in each spatial patch, and in a limited amount of time.
We denote by $f_\ct$ the frequency of GWs entering the Hubble radius when emission shuts off.
The remarkable property of the CT is that GWs with frequencies $f<f_\ct$ evolve independently of the source, because the corresponding wavelengths are larger than the source's correlation length.
The evolution of each tensor mode $h_k(t)$ in this regime is sensitive only to the expansion of the Universe and GW propagation \cite{
Seto:2003kc,
Weinberg:2003ur,
Boyle:2005se,
Watanabe:2006qe,
Boyle:2007zx,
Caprini:2009fx,
Jinno:2012xb,
Caprini:2015zlo,
Barenboim:2016mjm,
Geller:2018mwu,
Saikawa:2018rcs,
Cui:2018rwi,
Caldwell:2018giq,
Caprini:2018mtu,
Cai:2019cdl,
DEramo:2019tit,
Bernal:2019lpc,
Figueroa:2019paj,
Auclair:2019wcv,
Chang:2019mza,
Hajkarim:2019nbx,
Caprini:2019egz,
Gouttenoire:2019kij,
Gouttenoire:2019rtn,
Domenech:2019quo,
Guo:2020grp,
Ellis:2020nnr,
Blasi:2020wpy,
Domenech:2020kqm,
Allahverdi:2020bys,
Berger:2023pon%
}.
Cosmological PTs are a typical example \cite{%
Caprini:2018mtu, 
Caprini:2019egz, 
Hindmarsh:2020hop,
Athron:2023xlk%
}. 
In this case, bubble collisions, sound waves, and plasma turbulence act as causality-limited sources, each with its own finite correlation length.

The GW energy fraction is customarily defined as
\begin{equation}
\label{eq:OGW}
\OGW(f)\equiv
\frac{1}{\rho_\text{cr}} \frac{\d \rho_\gw(f)}{\d \ln f} 
\end{equation}
where we introduced the critical energy density today $\rho_\text{cr}=\rho_{\gamma,0}/\Omega_{\gamma,0} $ in Eq.~\eqref{eq:OGW} (where $\Omega_{\gamma,0} h^2= 2.47\times 10^{-5}$ is the SM radiation abundance today and $h\equiv H_0/(100~\text{km/s/Mpc})$ is the reduced Hubble constant).
$\OGW(f)$ exhibits a spectral shape, that is typically peaked at some frequency $\fp>f_\ct$. 
The CT of the spectrum behaves as $\OGW(f\lesssim f_\ct)\propto f^3$ in a universe filled by a perfect relativistic fluid with EoS $w(t)=p/\rho=\tfrac 13$, see e.g.~\cite{Caprini:2009fx}.

This tilt of the CT can be modified as the GW energy density $\rho_\gw(f)$ in the CT is determined by the Universe's expansion history and its effect on evolution of tensor modes \cite{Hook:2020phx}. After the GW source shuts off, the emitted super-Hubble modes freeze due to Hubble friction, remain practically constant until they re-enter the Hubble sphere, and then proceed with under-damped oscillations diluted as $1/a$.
If the equation of state parameter is constant at this epoch, the CT scales as
\begin{equation}
\frac{\d \rho_\gw(f)}{\d \ln f}\bigg|_{f<f_\ct} \propto f^{3+2\frac{3w-1}{3w+1}}.
\end{equation}

\begin{figure}[t]\centering
  \includegraphics[width=\columnwidth]{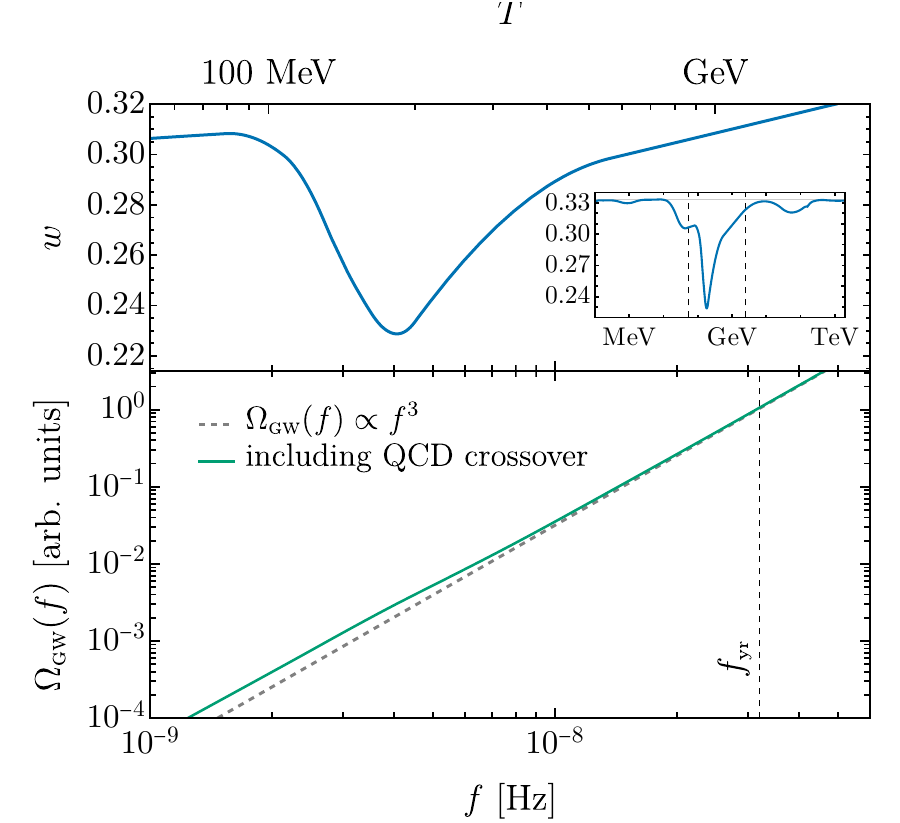}
  \vspace{-2em}
  \caption{
  \textit{Top:} EoS during the QCD crossover. The inset shows $w$ in an expanded range of temperatures.
  \textit{Bottom:} Impact of the variation of $w(T)$, $\gs(T)$ on the CT of a primordial SGWB (plotted with an arbitrary amplitude).
  The dashed line shows the $f^3$ scaling obtained in a pure radiation-dominated universe.}
  \label{fig:QCD-PT}
\end{figure}

On the other hand, during the QCD crossover the EoS is not constant, since heavy hadrons form and both the effective number of degrees of freedom in energy $g_{*}(T)=\rho/(\pi^2 T^4/30)$ and entropy densities $g_{*,\text{\tiny s}}(T)=s/(2\pi^2T^3/45)$ decrease rapidly. From the thermodynamical relation $sT=p+\rho$ one then infers the temperature-dependent EoS $w(T)=\tfrac 43(g_{*,\text{\tiny s}}(T)/g_{*}(T))-1$.
The values of $g_{*}(T), g_{*,\text{\tiny s}}(T)$, and the EoS shown in Fig.~\ref{fig:QCD-PT}, are precisely determined with lattice QCD techniques~\cite{Borsanyi:2016ksw}. The temporary decrease of $w(T)$ is due to the pressureless contribution of QCD matter to the thermal bath, followed by its rapid depletion.
The CT is thus distinctively modified, when the corresponding GWs with wavenumber $k$ re-enter the Hubble horizon during the QCD crossover. 
Their frequency is set by $f=k/2\pi=aH/2\pi$, which in terms of temperatures reads (see App.~\ref{app:f hor cross})
\begin{equation}
\label{fTp}
f \simeq 3.0 \nHz \cdot 
\parfrac{g_{*,\text{\tiny s}}(T)}{20}^{1/6} 
\parfrac{T}{150 \MeV}.
\end{equation}
The effect of the temperature dependence of $w(T)$ on the CT can be only approximately captured (see e.g. \cite{Watanabe:2006qe} in the context of inflationary GWs)by accounting in \eqref{eq:OGW} for the fact that SM radiation is slightly reheated during the crossover, such that:
\begin{equation}\label{Omrad}
\rho_\gamma(T)=
\rho_{\gamma,0} 
\left(\frac{g_{*,\text{\tiny s}}(T_0)}{g_{*,\text{\tiny s}}(T)}\right)^{4/3} 
\left(\frac{g_*(T)}{g_*(T_0)}\right) \left(\frac{a_0}{a}\right)^4 \,,
\end{equation}
where $a (a_0)$ is the scale factor (today), while GWs are decoupled from the thermal bath.

We perform a precise computation by solving the equations of motion for $h_k(t)$ (see  App.~\ref{app:tabulated CT}), accounting for the temperature dependence of $w$ and $g_{*}$, and obtain the spectrum shown in Fig.~\ref{fig:QCD-PT}. 
A clear deviation from the commonly employed $f^3$ approximation is found, 
accidentally located in the nHz range where PTA experiments are most sensitive.

Beside the SM effects presented above, a nHz CT signal probes the cosmic expansion history back to $\sim T_\QCD$. 
In particular, the possible presence of a fraction $\fFS\equiv \rho_\FS/\rho_\text{tot}$ of free-streaming species is easily accounted for, by multiplying the $\OGW$ spectrum shown in Fig.~\ref{fig:QCD-PT} by a factor $f^{16\fFS/5}$, for $\fFS\lesssim 15\%$~\cite{Hook:2020phx}.

%%%%%%%%%%%%%%%%%%%%%%%%%%%%%%%%%%%%%%%%%%%%%%%%%%%%%%%%%%%%%%%%%%%%%%%
\section{SGWBs at PTAs}
\label{sec:pta}
%%%%%%%%%%%%%%%%%%%%%%%%%%%%%%%%%%%%%%%%%%%%%%%%%%%%%%%%%%%%%%%%%%%%%%%

A SGWB is associated to a characteristic strain $h_c(f)\simeq 1.26\cdot 10^{-9}\cdot(\text{nHz}/f)\sqrt{\OGW(f) h^2}$ \cite{Caprini:2018mtu}.
This induces the cross-power spectral density for the timing residuals $S_{ab} (f)=\Gamma_{ab} h_c^2(f)/(12\pi^2 f^3)$, where $\Gamma_{ab}$ models correlations between pulsars $a$ and $b$ and is given by the Hellings-Downs function~\cite{Hellings:1983fr}.
PTA collaborations typically assume a power-law $h_c(f) = A_\cp(f/f_\yr)^\alpha$ ($S_{ab}(f)\propto (f/f_\yr)^{-\gamma}$ with $\gamma=3-2\alpha$), so that the corresponding GW fraction is 
\begin{equation}
\label{eq:pl}
\Omega_\cgw(f)h^2 \simeq 6.3\cdot 10^{-10}\left(\frac{A_{\cgw}}{10^{-15}}\right)^2\parfrac{f}{f_\yr}^{n_T},
\end{equation}
where $f_\yr\equiv (1~\yr)^{-1}\simeq 32~\nHz$ and $n_T=2(\alpha+1)=5-\gamma$. 
For a cosmological SGWB, we use the subscript $\cgw$ rather than $\cp$. While a SGWB from SMBHBs is expected, its properties are not yet fully known. In particular, population synthesis studies predict a wide range for the amplitude and $\alpha$ \cite{Haiman:2009te,Kocsis:2010xa, Burke-Spolaor:2018bvk, Middleton:2020asl,Pan:2021oob, Antoniadis:2022pcn,Ellis:2023owy,Kozhikkal:2023gkt, NG15-NP, EPTA2-SMBHB-NP}.
The simple prediction $n_T= \tfrac 23$ (i.e.~$\alpha=- \tfrac 23$, $\gamma=\tfrac{13}{3}$) is only valid for a continuous population with circular orbits and GW-driven energy loss~\cite{Phinney:2001di}.   

The causality tail (CT) of a \cgw{} is commonly modeled as a power-law with $n_T=3$ ($\alpha=\tfrac 12, \gamma=2$).
 The main novelty of our work is that the actual CT is significantly different, when $f_\ct$ lies above the frequency bins employed in PTA analyses, i.e.~$10\nHz\lesssim f_\ct \lesssim \fp$. A convenient parametrization of the CT signal is
\begin{equation}
\label{eq:hCT}
  h_{c,\ct}(f)= A_\ct~\frac{\S(f)}{\S(f_\ct) }
  \left(\frac{f}{f_\ct}\right)^{\frac{8}{5} \fFS},
\end{equation}
where $\S(f)$ is tabulated in App.~\ref{app:tabulated CT} and normalised to $\S(f_\yr)=1$, while $A_\ct$ is the amplitude at $f_\ct$. Fig.~\ref{fig:QCD-PT} shows $\OGW(f)\propto f^2 \S^2(f)$ (i.e. with $\fFS = 0)$. 

A crucial difference between astrophysical and \cgw{} backgrounds is that \textit{only} the latter contribute to the energy budget in the early Universe, and can affect cosmological observables as any other relativistic free-streaming component beyond the SM. 
\cgw{}s contribute to the {\it effective number of neutrino species} as $\Neff\equiv 3.044+\Delta N^{\cgw}_{\text{eff}}$, with $\Delta N^{\cgw}_{\text{eff}}=\rho_\cgw/\rho_{\nu,1}$ and $\rho_{\nu,1}$ is the energy density of a single neutrino species. 
Specifically, the total (integrated) GW abundance is $\Omega_\cgw h^2\simeq 1.6\cdot 10^{-6}\left(\Delta N^{\cgw}_{\text{eff}}/0.28\right)$~ \cite{Caprini:2018mtu}. 
Measurements of the Cosmic Microwave Background (CMB)~\cite{Planck:2018vyg} and Baryon Acoustic Oscillations (BAO) constrain $\Delta \Neff\leq 0.28$ at $95\%$ C.L. For the peaked sources of interest, $\Omega_\cgw\simeq \Omega_\cgw (\fp)$ and the constraint on \cgw{} backgrounds reads
\begin{equation}
\label{eq:neffbound}
A_\cgw\leq 5\times 10^{-14}\parfrac{f_\yr}{\fpk}^{1+\alpha}, \quad \ 
\text{(95\% C.L.)},
\end{equation}
for signals that can be approximated with power-laws up to $f_\star$. 
While this assumption is often not valid (see e.g.~\cite{Cutting:2020nla, Jinno:2020eqg} for PTs), the $\Delta \Neff^{\cgw}$ bound can be applied model-independently to the CT, thereby giving $A_\ct\leq 5\times 10^{-14}(f_\yr/f_\ct)$. 
This often captures the approximate strength of the constraint, since typically $f_\ct\lesssim (0.1-1)f_\star$ and the spectrum flattens close to the peak.
By comparison with Eq.~\eqref{eq:pl}, it is evident that CMB constraints can affect signal interpretation when $f_\ct\gtrsim f_\yr$, if $\alpha\geq -1$ (see also \cite{Moore:2021ibq}). 

Finally, let us remark that GWs with frequency $f\gtrsim f_\ct$ provide an unavoidable contribution to $\fFS \sim 0.3\Delta \Neff^\cgw$, see also App.~\ref{app:tabulated CT}.

%%%%%%%%%%%%%%%%%%%%%%%%%%%%%%%%%%%%%%%%%%%%%%%%%%%%%%%%%%%%%%%%%%%%%%%%
\section{Bayesian Analysis}
\label{sec:bayes}
%%%%%%%%%%%%%%%%%%%%%%%%%%%%%%%%%%%%%%%%%%%%%%%%%%%%%%%%%%%%%%%%%%%%%%

We now present the results of a Bayesian search for the CT of a \cgw{} signal in IPTA-DR2~\cite{Antoniadis:2022pcn} and in the recent NG15~\cite{NG15-SGWB} dataset.  
We aim to quantify the impact of the QCD-induced deviations from a power-law with $n_T=3$ on signal interpretation. 
Following the IPTA and NG15 collaborations, we limit our analysis to the first 13 and 14 frequency bins of each datasets respectively, to avoid pulsar-intrinsic excess noise at high frequencies. 

We consider two \cgw{} models: the CT and the power-law $\nt=3$, both with only one free parameter, $A_\ct$ and $A_\cgw$ respectively. We impose $\Delta \Neff$ constraints in (below) Eq.~\eqref{eq:neffbound} to $A_\cgw$ ($A_\ct$). 
For the CT signal, we fix $\fFS$ to the unavoidable contribution from GWs of frequency $f=f_\ct$ (see App.~\ref{app:tabulated CT} for details). We set $f_\ct=f_\yr$ (such that the first bins of both datasets are deep in the CT) and similarly $f_\star=f_\yr$ for the $n_T=3$ model, and comment on other choices below. 
We also compare with common power-law processes (non-\cgw) with free exponent $\gamma$, as well as with $\gamma = \tfrac{13}{3}$ (i.e. $n_T=\tfrac 23, \alpha=-\tfrac 23$). For the NG15 analysis, we also compare with the SMBHB expectation obtained by the NG15 collaboration~\cite{NG15-NP}, via population synthesis studies assuming circular orbits and GW-only energy loss. App.~\ref{app:priors} lists all parameters and prior choices.

We compare models by computing the Bayes factors $\mathcal B_{ij}$, quantifying the preference of model $i$ with respect to model $j$. 
These are reported in Tab.~\ref{tab:bayesPTA}. In this analysis, we consider only auto-correlation terms, rather than the full HD function, because of computational time limitations. We expect only minor effects in our NG15 results when including HD correlation~\cite{NG15-SGWB}.

\begin{table}[t!]
\centering
\begin{tabular} {ccccc}
\toprule
\multirow{2}{*}{\diagbox{PTA}{${\cal M}$}}
& ~~~$n_T = 3$~~~  & ~~Free~~ & ~~$\nt=\tfrac 23$~~ & SMBHB  \\
 & $(\gamma = 2)$ & $\nt$ & ($\gamma=\tfrac{13}{3}$) & (GW-driven) \\
\midrule
IPTA-DR2 &  \textbf{0.8} & $-2.3$ & $-3.2$ & --- \\
\midrule
NG15 & \textbf{1.0} & $-1.6$ & $-0.7$ & $-0.1$ \\
\bottomrule
\end{tabular}
\caption{$\log_{10}(\mathcal B_{\text{Causality Tail},\,\mathcal M})$ for the comparison of the CT with alternative models.
The CT spectrum (see Fig.~\ref{fig:QCD-PT}) is \textit{substantially} preferred with respect to $\OGW\sim f^3$.
The $\Delta \Neff$ upper bound has been imposed on the amplitude of both CT and $f^3$ spectra, fixing $f_\ct=f_\yr$ for the CT and $f_\star=f_\yr$ for $n_T=3$.}
\label{tab:bayesPTA}
\end{table}%

One important result is the comparison between the CT and power-law $n_T=3$ signals. 
As highlighted in bold in Tab.~\ref{tab:bayesPTA}, the CT provides a better fit than $f^3$ to both datasets, with a Bayes factor indicating ``substantial evidence". 

The comparison with a possible astrophysical SGWB depends on the dataset, and is subject to uncertainties on the SMBHB signal. 
For NG15, no substantial evidence in favor of nor against the CT is observed in comparing with the SMBHB expectation of~\cite{NG15-NP}, while the $n_T=\tfrac 23$ and the free power law models are both favored over the CT. The preference for these two power law models is stronger in the older IPTA-DR2.

\begin{figure}[t]\centering
\includegraphics[width=\columnwidth]{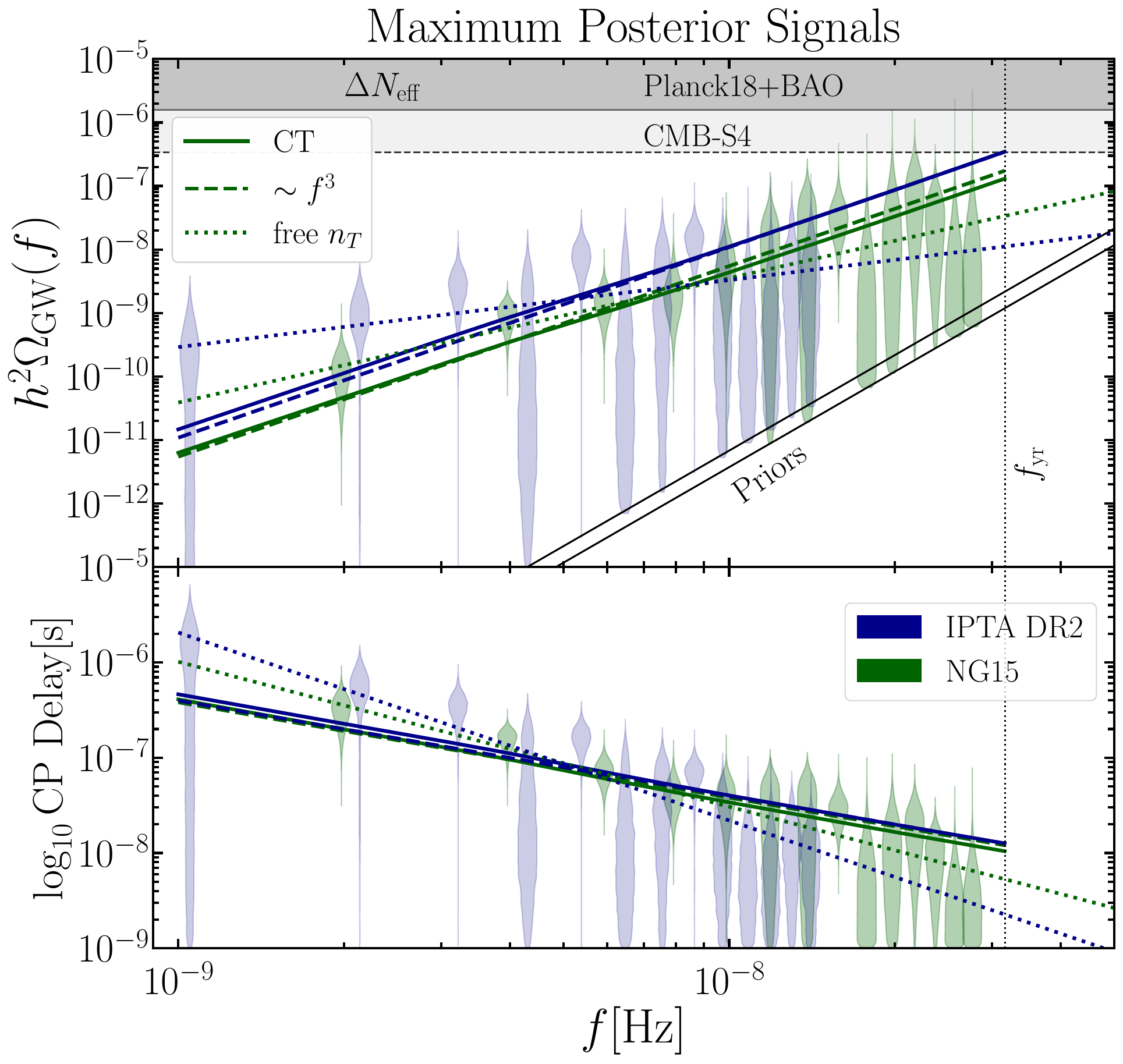}
\caption{
The GW spectrum (\textit{upper}) and the induced CP time delays $\sqrt{S_{ab}(f)/T}$  (\textit{lower}), $T$ being the observation time of each PTA.
We compare CT (solid), $n_T=3$ (dashed) and free power-law (dotted) models, selecting the maximum posterior values obtained with IPTA-DR2~\cite{Antoniadis:2022pcn} (blue) and NG15~\cite{NG15-SGWB} (green) datasets. The shaded region shows the CMB+BAO bound on $\Delta \Neff$, and the dotted line the reach of CMB-S4 experiments.
The lower limits of the posteriors are determined by the priors of~\cite{Antoniadis:2022pcn, NG15-SGWB}.
See App.~\ref{app:priors} for parameter posteriors.}
\label{fig:bestfit}
\end{figure}

The GW spectra and CP time delays for maximum-posterior values of parameters obtained by our analyses are shown in Fig.~\ref{fig:bestfit} (here we include HD correlations in the NG15 results, using the faster fitting procedure described in~\cite{Mitridate:2023oar}). Notice that for $A_\ct = A_\cgw$ the CT allows for a larger amplitude of $\OGW$ and of CP delays in the first bins (roughly by a factor of $1.5$).

The curves in Fig.~\ref{fig:bestfit} correspond to negligible values of $\fFS$ from GWs of frequency $f=f_\ct$. Additional contributions from high-frequency and other free-streaming species are model-dependent, but would affect our results only if $\Delta\Neff^\cgw\gtrsim 0.1$. In this case, an excess in upcoming cosmological surveys~\cite{DESI:2016fyo, Amendola:2016saw, CMB-S4:2022ght} is expected.

For a qualitative comparison with simple power-laws, one can approximate the CT as a power-law with slope evaluated through a given number of frequency bins. 
The result of this procedure is shown by the vertical gray lines in Fig.~\ref{fig:NG15}, obtained by fitting the CT with a power-law through the lowest 4th to 8th frequency bins of NG15, where evidence for GWs is reported (and within the 9 frequency bins employed by EPTA-DR2).  The $1,2$ and $3\sigma$ posteriors on power-law parameters reported by IPTA-DR2 and NG15 are also shown, together with the results of EPTA-DR2 (using the {\tt DR2new HD} dataset). 
We expect the CT to provide as good a fit to the NG15 (EPTA-DR2) data as a power-law model with parameters inside the $3\sigma$ ($1\sigma$) region of the posteriors, whereas the $f^3$ signal lies at the border of the $3\sigma$ ($2\sigma$) region. 
We stress that this is only approximate, since the CT  deviate significantly from a power-law at the relevant frequencies. 

Let us comment on how different $f_\ct$ values affect our model comparison.  
The CMB bound on power-law $\cgw$s becomes stronger for $f_\ct> f_{\yr}$, as shown in Fig.~\ref{fig:NG15} for peak frequency $\fpk=1,3,10f_\yr$. Thus, we find that the IPTA-DR2 dataset decisively disfavors $f_\ct\gtrsim 32 \nHz$, while for the new datasets the CT signal is excluded at $3\sigma$ (affected) only if $f_\ct\gtrsim 100 \nHz$ ($\gtrsim 50 \nHz$).
The future reach of CMB-S4~\cite{CMB-S4:2022ght} observations is also plotted.

%%%%%%%%%%%%%%%%%%%%%%%%%%%%%%%%%%%%%%%%%%%%%%%%%%%%%%%%%%%%%%%%%%%%%%%%
\section{Discussion and Implication for Particle Physics}
\label{sec:discussion}
%%%%%%%%%%%%%%%%%%%%%%%%%%%%%%%%%%%%%%%%%%%%%%%%%%%%%%%%%%%%%%%%%%%%%%%

\begin{figure}[t]\centering
  \includegraphics[width=\columnwidth]{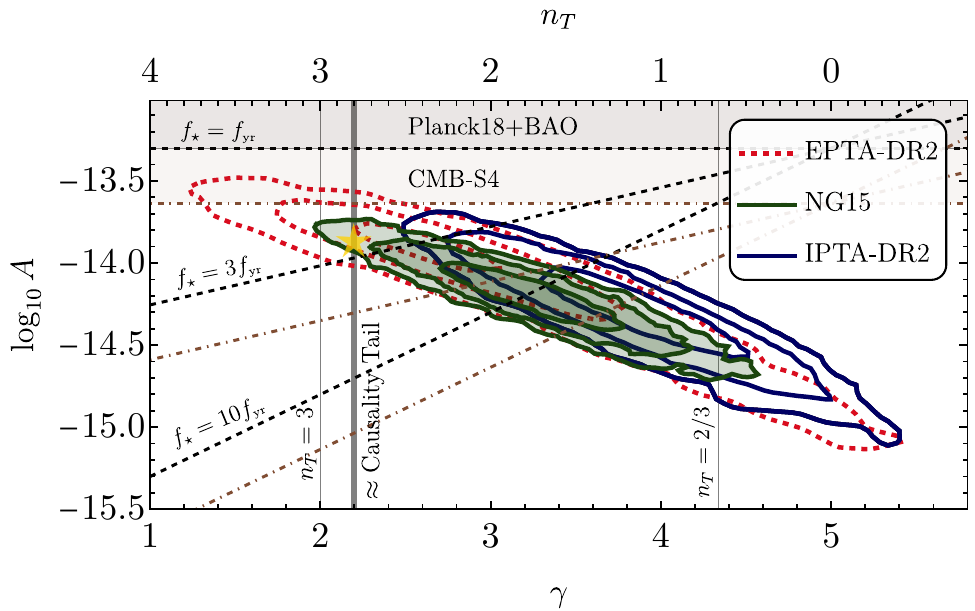}
  \vspace{-2em}
  \caption{
  IPTA-DR2~\cite{Antoniadis:2022pcn}, NG15 and EPTA-DR2
  posteriors for a power-law model, the last two enforcing HD correlations in the analyses~\cite{NG15-SGWB, EPTA2-SGWB}. 
  The vertical lines indicate approximations of the CT.
  The yellow star shows the best fitting CT amplitude according to NG15 data.
  We show current and future $\Delta \Neff$ bounds affecting any power-law signal extending up to $\fpk = 1,3,10\, f_\yr$.}
  \label{fig:NG15}
\end{figure}

We have pointed out that SM physics predicts a specific shape for the CT of a GW signal, clearly distinguishable from $\OGW\sim f^3$ already with present datasets. 
This applies to any GW source that is active before QCD confinement (i.e.~$T\gtrsim \GeV$, see App.~\ref{app:particlephysics}), and provides a much-needed signature to help determining whether the current PTA excess is of cosmological or astrophysical origin.
If the SMBHB interpretation (or a different \cgw{} signal) is preferred, then our work should prove useful in the future to disentangle additional contributions from cosmological sources (see e.g.~\cite{Kaiser:2022cma,Sato-Polito:2023spo}).

We have quantified that, despite the CMB bound on $\Delta\Neff$~\cite{Moore:2021ibq}, a CT spectrum can explain the PTA signal and is substantially favored over the commonly employed $n_T=3$ approximation in both IPTA-DR2 and NG15 datasets, while providing a similar fit to NG15 as simple SMBHB models.

The detection of such a signal would have dramatic implications for particle physics, most likely pointing to the breaking of a (global or gauge) symmetry in a dark sector (DS), for instance via a PT (or causing the annihilation of domain walls for discrete symmetries) at temperatures $T\gtrsim\GeV$. 
Some macroscopic properties of the corresponding GW source can be estimated rather model-independently (see App.~\ref{app:particlephysics}). 
Denoting by $\alpha_{\star} \equiv [\rho_{\text{\tiny DS}}/(3 H^2 M_p^2)]_{T=T_\star}$ the energy fraction of the source at $T_\star$, and by $\epsilon_\star\lesssim 1$ the model-dependent suppression (e.g.~the sub-Hubble size of the source, see App.~\ref{app:particlephysics}), the posteriors ($2\sigma$) in Fig.~\ref{fig:NG15} imply the lower bound $\alpha_\star\gtrsim (0.1-0.4)\epsilon_\star^{-1/2}$, for $15 \nHz\lesssim f_\ct\lesssim f_\yr$.
The required fraction is even larger for larger $f_\ct$. 
This may be problematic for sources with $\epsilon_\star\ll 1$, as the DS would then dominate over the SM background ($\alpha_\star \simeq 1$), and thus suggests that the source should be relativistic and Hubble-sized. 
We discuss further the PT scenario and its challenges in App.~\ref{app:particlephysics} (see also \cite{Ferreira:2022zzo, Bai:2021ibt, Bringmann:2023opz}).

Such a large abundance should decay to the visible sector (or be in the SM, if e.g.~QCD confinement proceeds via a first order PT in the early Universe, although this scenario requires additional ingredients, see e.g.~\cite{Bodeker:2020stj}), because if it remains in dark relativistic species it contributes as~\cite{Ferreira:2022zzo}
\begin{equation}
\Delta \Neff^{\text{DS}}\simeq 0.28 \, 
\parfrac{74}{g_*(T_\star)}^{1/3} \parfrac{\alpha_\star}{0.1},
\end{equation}
where we normalised to the current $2\sigma$ constraints from CMB. Interestingly, the required couplings between the SM and the dark sector may be additionally probed at colliders, laboratory experiments and astrophysical environments. 
For non-CT $\cgw$ signals, NG15 gives a $2\sigma$ bound $A_\cgw\gtrsim 10^{-14.6}$ to explain the current excess, thus requiring $\alpha_\star\gtrsim 0.07\epsilon_\star^{-1/2}$, well within the reach of
upcoming CMB and LSS surveys~\cite{DESI:2016fyo, Amendola:2016saw, SimonsObservatory:2018koc, CMB-S4:2022ght}.

Finally, in our work we have conservatively employed the SM prediction for the EoS of the Universe. 
We notice that the measurement of the CT at PTAs offers a probe of the whole cosmological expansion history up to $T_\QCD$, as encapsulated by Eq.~\eqref{fTp}.
Modifications to the CT  could occur in BSM scenarios where $w(T)$, $g_*(T)$ or $\fFS$ are varied,
or if the Universe undergoes a phase of matter domination below the QCD crossover (we present a search for this scenario in App.~\ref{app:MD}). 
Additionally, PTAs could test the EoS of hot QCD matter in the early Universe (see e.g.~\cite{Bodeker:2020stj, Achenbach:2023pba}), in the presence of a CT signal.

The near-future detection of a SGWB at PTAs could disclose a unique window around the epoch of the QCD crossover. As highlighted in this paper, this coincidence of scales provides a robust feature to assist in the discrimination between the astrophysical and cosmological origin of the SGWB.

%\acknowledgments
\begin{acknowledgments}
We thank Peter Graham, Antonio Junior Iovino, Ville Vaskonen and Hardi Veermae for discussions, Zoltan Haiman for useful correspondence, and Valerie Domcke, Anson Hook and Gustavo Marques-Tavares for comments on the draft. We thank Rudin Petrossian-Byrne and Mehrdad Mirbabayi for discussions on our work.
We acknowledge use of the publicly available codes {\tt enterprise}~\cite{enterprise}, {\tt en\-ter\-pri\-se\_extensions}~\cite{enterprise_ext}, {\tt PTMCMC}~\cite{justin_ellis_2017_1037579}, {\tt PTArcade}~\cite{Mitridate:2023oar} and {\tt ceffyl}~\cite{Lamb:2023jls} as well as of the free spectrum chains publicly provided by the \href{https://zenodo.org/record/5787557}{IPTA-DR2} collaboration.
G.F. acknowledges financial support provided under the European
Union's H2020 ERC, Starting Grant agreement no.~DarkGRA--757480 and under the MIUR PRIN programme, and support from the Amaldi Research Center funded by the MIUR program ``Dipartimento di Eccellenza" (CUP:~B81I18001170001).
This work was supported by the EU Horizon 2020 Research and Innovation Programme under the Marie Sklodowska-Curie Grant Agreement No. 101007855 and 
and  additional financial support provided by ``Progetti per Avvio alla Ricerca - Tipo 2", protocol number
AR2221816C515921.
D.R.~is supported in part by NSF Grant PHY-2014215, DOE HEP QuantISED award \#100495, and the Gordon and Betty Moore Foundation Grant GBMF7946. The work of F.R.~is partly supported by the grant RYC2021-031105-I from the Ministerio de Ciencia e Innovación (Spain). F.R. thanks the Galileo Galilei Institute in Florence and the organizers of the New Physics@Korea Institute for kind hospitality during the completion of this work.
\end{acknowledgments}

%%%%%%%%%%%%%%%%%%%%%%%%%%%%%%%%%%%%%%%%%%

\bibliographystyle{apsrev4-2}
\bibliography{bib_SGWB-PTA.bib}

\appendix

%%% da 2 colonne a 1, con titolo
\clearpage
\onecolumngrid
{\large 
\centering
\textbf{APPENDIX}

\smallskip}

%%%%%%%%%%%%%%%%%%%%%%%%%%%%%%%
\section{Relation between frequency today and temperature at horizon crossing}
\label{app:f hor cross}
%%%%%%%%%%%%%%%%%%%%%%%%%%%%%%%

We derive here the present frequencies corresponding to modes crossing the horizon at a given time, in order to clarify a potential confusion with the typical peak frequency of primodial SGWBs and to highlight the coincidence of frequencies probed by PTAs and modes corresponding to the QCD crossover. 

The frequency (redshifted to its present day value) of modes crossing the horizon at a given epoch can be simply derived as follows. We impose that, at the temperature of horizon crossing one has $k = a H$, 
\begin{equation}
f(T) \equiv \frac{k}{2 \pi a_0} = 
  \frac{a(T)}{ a_0}\frac{ H(T) }{2 \pi}.
\end{equation}
Using the conservation of entropy 
$g_{*,\text{\tiny s}}(T)T^3a^3(T) =$ constant,
the ratio of scale factors between GW production at a temperature $T$, and the present time at $T_0$, is
\begin{equation}
\label{a0ap}
\frac{a(T)}{a_0} = 
\left(\frac{g_{*,\text{\tiny s}}(T_0)}{g_{*,\text{\tiny s}}(T)}\right)^{1/3} \parfrac{T_0}{T} \simeq 
8.0 \times 10^{-14} 
\parfrac{g_{*,\text{\tiny s}}(T)}{100}^{-1/3} \parfrac{T}{\GeV}^{-1},
\end{equation}
where we have used $T_0 \simeq 2.4\times 10^{-13}$ GeV for the photon temperature today \cite{ParticleDataGroup:2016lqr} and $g_{*,\text{\tiny s}}(T_0) \simeq 3.91$ for the Standard Model degrees of freedom with three light neutrino species~\cite{Kolb:1990vq}. 
Additionally, employing the Friedmann equation $H^2 = \rho /(3 \MP^2)$, where $\MP\equiv \sqrt{8\pi G_{N}}\simeq 2.4\cdot10^{18}~\text{GeV}$ is the reduced Planck mass, the energy density in a radiation dominated era $\rho = (\pi^2/30) g_*(T)T^4$ and $g_{*,\text{\tiny s}}(T) \simeq g_*(T)$, one obtains Eq.~\eqref{fTp}.

If one considers the emission of GWs from a cosmological phase transition (or from e.g. domain wall annihilation), the characteristic frequency $\fp$ of the SGWB, i.e. the frequency at which one expects the SGWB to peak, is located around the inverse typical time or length scale of the system, which is the inverse of the duration of the PT or of the bubble size, depending on the details of the source.
Following Ref.~\cite{Caprini:2018mtu}, we set $k_\star/a_\star\simeq 2\pi \,\beta$, where $\beta$ is the inverse of the nucleation timescale. Therefore, one obtains the characteristic frequency today as
\begin{equation}
\label{eq:pfreq}
f_{\star,0} \simeq 24 \nHz \cdot \parfrac{\beta}{H_\star} \parfrac{g_*(T_\star)}{100}^{1/6}\parfrac{T_\star}{150\MeV}.
\end{equation}
Notice that this frequency is larger by a factor of $2\pi$ than the frequency in Eq.~\eqref{fTp} corresponding to horizon crossing at a given $T$, because they refer to two different physical quantities. 
Eq.~\eqref{eq:pfreq} gives the typical peak $\fp$ of the SGWB and identifies the wavenumber $k = 2\pi /\lambda = 2\pi f$ associated to the excitation of a GW over a period $f^{-1}\simeq\beta^{-1}$, whereas Eq.~\eqref{fTp} determines the wavenumber $k=a(T)H(T)$ crossing the Hubble radius at a given time, and allows to fix the boundary $f_\ct$ of the CT.

%%%%%%%%%%%%%%%%%%%%%%%%%%%%%%%%%%%%%%%%%%%%%%%%%%%%
\section{Tabulated causality tail}
\label{app:tabulated CT}
%%%%%%%%%%%%%%%%%%%%%%%%%%%%%%%%%%%%%%%%%%%%%%%%%%%%
%

This Appendix provides more details about the CT that we compute and use for our analysis.
For completeness, we provide a summary of the dynamics for low frequency GW modes excited by a local, causality-limited source in the early Universe (see \cite{Hook:2020phx} for details).
The equations of motion in conformal time $\tau$ for a GW mode $h_k(\tau)$ sourced by sector with energy density $\rho_\ds$ and dimensionless anisotropic stress $\Pi$ read (we understand the space-time tensor indices and the polarisation $(+,\times)$ indices for both the GW and source)
\begin{equation}
h_k''+2 \mathcal H \,h'_k + \frac{k^2}{a^2} h_k
=a^2 \frac{4\rho_\ds}{3\MP^2} \Pi(\tau,k) \equiv J(\tau, k) \,,
\label{eq:GW EOM}
\end{equation}
where $\mathcal H=a'/a$ is the conformal Hubble time and $'$ denotes conformal time derivatives.
At small $k$, the source takes a simple and universal form when it lasts for a finite amount of time (up to a time $\tau_\star$) and has a finite correlation length ($\lesssim 2\pi / k_\star$, where $k_\star\equiv \mathcal H(\tau_\star)$ is the mode crossing the Hubble radius at $\tau_\star$).
This corresponds to GW periods longer than the duration of the source, so that the time dependence of the source is basically a Dirac delta function, and to wavenumbers where the power spectrum of the source is independent of $k$ (and is given by white noise, meaning that sources in different Hubble patches are uncorrelated). 
Therefore, in this limit $J(\tau,k)\to J_\star \delta(\tau-\tau_\star)$, so that an initially vanishing GW mode gets a velocity $h'_k=J_\star$ shortly after $\tau_\star$, and thereafter evolves according to 
\begin{equation}
\begin{gathered}
h''_k(\tau) + 2 \mathcal H(\tau) h'_k(\tau) + \frac{k^2}{a^2(\tau)} h_k(\tau) =0 \,,\\
h_k(\tau_\star)=0\,, \quad h'_k(\tau_\star)=J_\star \,.
\label{eq:GW EOM CT}
\end{gathered}
\end{equation}
We solve Eq.~\eqref{eq:GW EOM CT} for a wide range of momenta $k<k_\star$ by accounting for the expansion history of $\Lambda$CDM with the SM content around the QCD crossover. 
We notice that a positive detection of these features on the CT allows to confirm that the cosmological expansion history up to $T\sim T_\QCD$ has followed the predictions of $\Lambda$CDM, and possible deviations would hint at a different evolution of the scale factor.
We can then compute the GW power spectrum $P_h\sim\langle(a(\tau) h_k(\tau))^2\rangle$ after horizon crossing, and $\Omega_\ct \propto f^5 P_h(f)$
\cite{Hook:2020phx}.

%%%%%%%%%%%%%%%%%%%%%%%%%%%%%%%%%%%%%%%

To include the effect of free-streaming (high-frequency) GWs on the CT, one starts with the modification on $\Omega_\gw h^2= S_1(f) f^{16 f_\text{FS} /5}$. Here $S_1$ is a function that describes the CT for $\fFS=0$. Then we transform to the characteristic strain $h_c(f)$, as follows
\begin{equation}
h_c(f)= 1.26\cdot 10^{-9}\left(\frac{\text{nHz}}{f}\right)\sqrt{\Omega_\gw h^2},
\end{equation}
which can then be rewritten as
\begin{equation}
\label{eq:hcf}
h_c(f)= A_\ct\frac{\S(f)}{\S(f_\ct)}\parfrac{f_\ct}{f}^{-\tfrac 85 \fFS},
\end{equation}
where $\S$ is the function tabulated in Tab.~\ref{table:tabulated}. 
When $f_\ct=f_{\text{yr}}$, then $A_\ct$ is the amplitude of the signal at $f=\text{yr}^{-1}$, as in our paper, whereas generally it is just the amplitude at $f_\ct$. With this notation, the $\Delta \Neff$ constraint on $\Omega_\gw h^2$ at $f_\ct$ reads
\begin{gather}
\OGW(f_\ct)h^2\simeq 1.6\cdot 10^{-6}\left(\frac{\Delta \Neff^\cgw}{0.28}\right) ,
\end{gather}
which implies
\begin{gather}
A_\ct= 5\cdot 10^{-14}\left(\frac{f_\yr}{f_\ct}\right)\sqrt{\frac{\Delta \Neff}{0.28}}.
\end{gather}
Now let us relate $\fFS$ to $A_\ct$. 
The definition of $\Delta \Neff$ is
\begin{equation}
\Delta \Neff=\frac{\rho_\gw}{\rho_{\nu,1}}\bigg|_\textsc{cmb},
\end{equation}
where $\rho_{\nu,1}$ is the energy density in one neutrino species. At CMB, this is related to the total energy density in radiation by
\begin{equation}
\rho_{\nu,1}= \rho_\text{rad}\left(\frac{T_{\nu}}{T_{\text{rad}}}\right)^4\left(\frac{g_\nu}{g_\text{rad}}\right)=
\rho_{\text{rad}}\left(\frac{4}{11}\right)^{4/3}\frac{2\cdot \tfrac78}{
g_{\star,\textsc{cmb}}
}. 
\end{equation}
Additionally, $\rho_\text{GW}/\rho_\text{rad}\propto g_{\star}\,g_{\star,\text{\tiny{s}}}^{-4/3}$ from entropy conservation, therefore we have
\begin{gather}
\Delta \Neff=\left(\frac{11}{4}g_{\star\text{\tiny s},\textsc{cmb}} \right)^{\frac 43}
\frac{ g_\star (T_\ct)^{1/3} }{ 2\cdot\tfrac 78 } 
 \cdot \fFS%\\
\simeq 13.7~g^{-1/3}_\star \fFS,
\end{gather}
where $\fFS= \rho_\gw/\rho_\text{rad}$ and we assume that the background is always dominated by SM radiation between GW production and matter-radiation equality. Using the above relation between $\Delta \Neff$ and $\OGW h^2$, we then have
\begin{equation}
\fFS\simeq 
0.086
\left(\frac{A_\ct}{5\cdot 10^{-14}}\right)^2\left(\frac{g_\star(T_\ct)}{74}\right)^{1/3},
\end{equation}
where we have normalized $g_\star$ to its value at temperatures corresponding to $f_\ct=f_{\text{yr}}$. Plugging this into \eqref{eq:hcf}, we obtain the GW-altered causality-tail characteristic strain as a function of two parameters: $f_\ct$ and $A_\ct$. In our runs, we shall fix $f_\ct$ and we are thus left with only one free parameter.

\begin{table}[t]\centering\begin{small}
\begin{tabular} {ccc}
\toprule
~~$f\, [\nHz]$~~ & ~~$\S(f)$ (Eq.~\ref{eq:hCT})~~ & ~~$\Omega_\ct(f)$ (arb. units)~~\\
\midrule
$1.$ & $0.229$ & $5.24 \cdot 10^{-5}$ \\ 
$1.19$ & $0.247$ & $8.56 \cdot 10^{-5}$ \\ 
$1.41$ & $0.267$ & $1.4 \cdot 10^{-4}$ \\ 
$1.67$ & $0.287$ & $2.28 \cdot 10^{-4}$ \\ 
$1.97$ & $0.309$ & $3.72 \cdot 10^{-4}$ \\ 
$2.34$ & $0.334$ & $6.07 \cdot 10^{-4}$ \\ 
$2.77$ & $0.361$ & $9.98 \cdot 10^{-4}$ \\ 
$3.29$ & $0.389$ & $0.00163$ \\ 
$3.9$ & $0.417$ & $0.00263$ \\ 
$4.62$ & $0.442$ & $0.00416$ \\ 
$5.48$ & $0.467$ & $0.00653$ \\ 
$6.49$ & $0.495$ & $0.0103$ \\ 
$7.7$ & $0.527$ & $0.0164$ \\ 
$9.13$ & $0.564$ & $0.0264$ \\ 
$10.8$ & $0.607$ & $0.043$ \\ 
$12.8$ & $0.656$ & $0.0704$ \\ 
$15.2$ & $0.708$ & $0.115$ \\ 
$18.$ & $0.765$ & $0.189$ \\ 
$21.4$ & $0.828$ & $0.312$ \\ 
$25.3$ & $0.898$ & $0.514$ \\ 
$30.$ & $0.974$ & $0.851$ \\ 
$35.6$ & $1.06$ & $1.41$ \\ 
$42.2$ & $1.15$ & $2.34$ \\ 
$50.$ & $1.25$ & $3.92$ \\ 
\bottomrule
\end{tabular} %
\caption{Tabulated values for the Causality Tail computed in this work, accounting for the effect of the SM QCD phase transition. The last column assume negligible $f_\text{\tiny FS}$.}
  \label{table:tabulated}
\end{small}
\end{table}

Table~\ref{table:tabulated} contains the tabulated result, parameterised through the function $\S(f)$ defined in Eq.~\eqref{eq:hCT}, which expresses the characteristic strain as a function of frequency in the PTA range.
This function can be directly used as an input for Bayesian parameter estimation.
We also include the tabulated curve for the corresponding $\Omega_\ct(f)\propto f^2 \S(f)^2$ shown in Fig.~\ref{fig:QCD-PT}.

%%%%%%%%%%%%%%%%%%%%%%%%%%%%%%%%%%%%
\section{Implications for Particle Physics}
\label{app:particlephysics}
%%%%%%%%%%%%%%%%%%%%%%%%%%%%%%%%%%%%
We collect here various important considerations from cosmology and particle physics related to our work, and in particular some consequences of the microscopical interpretation of the GW primordial signal.
We focus especially on the SGWB generated by the prototypical example of a causality-limited signal, a first-order phase transition.

%%%%%%%%%%%%%%%%%%%%%%%%%%%%%%%
\paragraph{Peak frequency for a causality-limited GW signal around the QCD phase transition.}
%%%%%%%%%%%%%%%%%%%%%%%%%%%%%%%
In this work we are studying the impact of the QCD era%
\footnote{The lattice results for $g_{*,\text{\tiny s}}(T)$ are obtained at vanishing baryon chemical potential, which corresponds to a vanishing baryon asymmetry at the QCD epoch, $\eta_B\ll 1$. Larger values are indeed expected to affect the EoS, see e.g.~\cite{Borsanyi:2021sxv}, therefore the CT. While still possible in principle, this scenario likely requires a late entropy injection that would also dilute any GW signal.}
 on modes 
that are excited by a local GW source when they are still outside the Hubble horizon, 
i.e.~$k_\QCD \ll a (T_\QCD) H (T_\QCD)$, where we denote by the subscript ``QCD'' quantities evaluated when the temperature of the universe was $T = T_\QCD \simeq 150 \MeV$.
The mode $k_\QCD$ corresponds to frequencies 
$f_\QCD \equiv k_\QCD/(2 \pi) \ll a H /(2 \pi)$.
Therefore, for consistency, we have to require the emission of frequencies around $f_\QCD\approx 3\nHz$ to take place when the corresponding modes were super-horizon, i.e. when the temperature of the universe was larger than $T_\QCD$.
Requiring $f_\QCD$ to be emitted at temperature larger than the QCD transition, forces the peak frequency to 
$\fp \gtrsim 24 \nHz \left(\beta/H_\star\right)$, neglecting the small impact of the change of effective degrees of freedom. 
Hence, a detection of a CT signal at PTAs would necessarily imply $T_\star\gtrsim 300\MeV$, or the peak of the signal would lie in the first PTA bins.
Future PTA measurements, while rapidly decreasing their sensitivity at frequencies below the inverse of the observation time \cite{Hazboun:2019vhv,DeRocco:2022irl,DeRocco:2023qae}, will significantly improve their sensitivity at frequencies larger than the presently constraining bins, thanks to the expected improvement in the instrumental noise in detectors like FAST \cite{Jiang:2019rnj} and MeerKat \cite{Miles:2022lkg}.

The possibility of a PT occurring in a DS not far from the SM QCD crossover might be motivated for example in proposals addressing the approximate coincidence of dark matter and baryonic energy density in the context of asymmetric dark matter \cite{Kaplan:2009ag,
Foot:2003jt,Foot:2004pq,Lonsdale:2014wwa,Lonsdale:2014yua,Lonsdale:2018xwd,Murgui:2021eqf}.

%%%%%%%%%%%%%%%%%%%%%%%%%%%%%%%
\paragraph{Constraints on the evolution of the dark sector responsible for the SGWB.}
%%%%%%%%%%%%%%%%%%%%%%%%%%%%%%%
The GW signal investigated in our work is largely independent of the nature of the GW source, as long as the latter occurs on sub-horizon scales and is active only up to a definite epoch in the early Universe. While this is obviously a virtue, it also implies that the detection of a CT signal at PTAs would not be able to uniquely identify the microphysical mechanism which acts as a GW source. To this aim, complementary probes would then be essential.

Nonetheless, some general conclusions can be drawn from a detection of a CT, which apply to a broad class of sources. These necessarily involve physics beyond the SM, which can be organized in a non-specified dark sector (DS). 
Several scenarios of interest can then be described by using effective macroscopic parameters: the temperature $T_\star$ at which the GW source is active and the fraction of the (radiation) energy density that it comprises, $\alpha_{\star} \equiv [\rho_{\text{\tiny DS}}/(3 H^2 M_p^2)]_{T=T_\star}$. 
The peak frequency of the GW signal at the time of emission is then determined by the inverse characteristic time scale of the source in units of the Hubble rate: this is given in Eq.~\eqref{eq:pfreq}, where we can define in greater generality a dimensionless parameter $\delta_\star \equiv \fp/H_\star$ (that takes the specific form $\beta/H_\star$ for PTs, while for domain walls in the scaling regime $\delta_\star\simeq 1$).
Typically, this time scale is smaller than or comparable to one Hubble time, thus $\delta_\star \gtrsim 1$. 
The peak amplitude of the signal today can be estimated as~\cite{Caprini:2018mtu}
\begin{equation}
\label{eq:omegapeak}
\Omega_\cgw(f_{\star,0})h^2\simeq
10^{-8}\, \epsilon_\star
\parfrac{74}{g_*(T_\star)}^{1/3} \parfrac{\alpha_\star/(1+\alpha_\star)}{0.14}^2,
\end{equation}
where  $g_*(T_\star)$ is normalized to $T_\star\simeq \GeV$, and $\epsilon_\star\lesssim 1$ generically scales as an inverse power of $\delta_\star$ (depending on the microphysics of the source) and also accounts for additional signal-suppressing effects (such as subluminal velocities).
For simple power-law signals below the peak frequency $\fp$, combining with Eq.~\eqref{eq:pl} one has
\begin{equation}
\alpha_\star\simeq 0.35~\epsilon_\star^{-\frac{1}{2}}
\parfrac{g_*(T_\star)}{74}^{\frac{1}{6}}\parfrac{A_\cgw}{10^{-14}}
\parfrac{\fp}{f_\yr}^{\nt/2},
\end{equation}
for $\alpha_\star\ll 1$. 
This estimate yields a lower bound on $\alpha_\star$ for CT signals, by replacing $\fp\rightarrow f_\ct, A_\cgw\rightarrow A_\ct$. 
In particular, for the $2\sigma$ posteriors shown in Fig.~\ref{fig:NG15} lead to $\alpha_\star\gtrsim (0.1-0.4)\epsilon_\star^{-1/2}$, for $15 \nHz\lesssim f_\ct\lesssim f_\yr$.
For larger values of $f_\ct$, this bound becomes even stronger. 
As mentioned in the main text, the required large fraction of energy density to fit the current excess cannot remain secluded from the SM bath below $T_\star$ because of CMB and BBN constraints. 
This leaves only one possibility: the dark sector should deposit a significant amount of its energy density into the SM bath shortly after the emission of GWs. 

This conclusion generically remains true when the source is a first order PT in a dark sector, even when the peak frequency lies among the first frequency bins of PTAs. Indeed, for PTs $\delta_\star\simeq \left(\beta/H_\star\right)\geq 1$ is the rate of variation of the bubble nucleation rate and is typically $\mathcal O(100)$ (or possibly smaller, depending on the exact time-dependence of thermal effective potential), while $\epsilon_\star\sim \delta_\star^{-2(-1)}v^{3(1)}_w$ if the signal is sourced by bubble collisions (sound waves), where $v_w\leq 1$ is the bubble wall velocity (see e.g.~\cite{Caprini:2018mtu}). 
Therefore, achieving such a large GW signal from a PT at a scale not far from 100 MeV requires a large value of $\alpha_\star$,%
\footnote{For PTs, $\alpha_\star\gtrsim 1$ implies that the PT is supercooled and the Universe undergoes a low-scale inflationary phase. A viable cosmology then requires exiting such a supercooled phase and reheating the Universe above Big Bang Nucleosynthesis, which might be challenging in concrete models, see e.g.~\cite{Baratella:2018pxi} for a discussion in a different context. A similar remark applies to the case of domain walls (DWs), although in this case exit from DW domination is likely not possible, thus $\alpha_\star\leq 1$ should be imposed.} 
and a mildly efficient damping of the DS energy density into the SM thermal bath before BBN. 
Such a coupling between the SM and new particles of $\mathcal O(\GeV)$ mass could lie in a wide range, as it would not suffer from significant constraints from supernova emission for such masses \cite{DeRocco:2019jti}, and could be probed in principle by laboratory experiments. 

A scenario that can potentially evade the constraints above, while still exhibiting a CT, is that of GWs induced by the collapse of large density fluctuations \cite{Acquaviva:2002ud, Mollerach:2003nq, Baumann:2007zm,Espinosa:2018eve, Kohri:2018awv, Domenech:2019quo}.
This can occur if the curvature power spectrum is enhanced at scales $k\gtrsim 10^6~\text{Mpc}^{-1}$, much smaller than those probed by CMB anisotropies \cite{Saito:2008jc,Vaskonen:2020lbd,DeLuca:2020agl,Bian:2020urb,Kohri:2020qqd,Sugiyama:2020roc,Domenech:2020ers,Inomata:2020xad,Franciolini:2022pav,Franciolini:2022tfm,Dandoy:2023jot,Ferrante:2023bgz}. 
However, for the CT to be relevant in the induced GW scenario, the peak of the SGWB should be reached at frequencies higher then those observed with PTA experiments, 
and with a peak amplitude $h^2\OGW(\fp)\gtrsim 10^{-8}$ (see also \cite{Dandoy:2023jot}).
Therefore, the required large scalar perturbation amplitude would generate an unacceptably large relic abundance of PBHs (see e.g. \cite{Sasaki:2018dmp,Carr:2020gox} for reviews), unless negative non-Gaussianities suppressing PBH formation are present \cite{FIVV} (overproduction of PBHs in Gaussian models has been shown to significantly constrain the interpretation of the excess already in the IPTA-DR2 and NANOgrav 12.5 years datasets in Ref.~\cite{Dandoy:2023jot}). 
Thus, a viable explanation of the PTA signal by second order GW requires the scalar source to be active around the horizon crossing of nHz modes, with a different specific imprint from the QCD crossover era (see e.g. \cite{Hajkarim:2019nbx,Abe:2020sqb}).

Finally, let us stress that our discussion does not generically apply to sources that are active on time scales much larger than one Hubble time, such as cosmic strings. 
These sources are not expected to exhibit a CT, and their GW signal is also not characterized by Eqs.~\eqref{eq:pfreq} and \eqref{eq:omegapeak}.
However, the interpretation of the PTA excess in terms of strings generated by the spontaneous breaking of a global symmetry (such as arise in axion models, notice that if the axion is massive then the string network disappears once the corresponding domain walls form and the GW signal then features a CT as usual) faces constraints from cosmology, see~\cite{Gorghetto:2021fsn}.

%%%%%%%%%%%%%%%%%%%%%%%%%%%%%%%
\paragraph{Impact of the bound from $\Delta \Neff$ on the amplitude of $\OGW(\fp)$.}
%%%%%%%%%%%%%%%%%%%%%%%%%%%%%%%
This consistency condition, that only applies to the scenario where the source for GW emission is a cosmological phase transition, becomes relevant when confronted with the requirement of falling below the bound from $\Delta \Neff$.
Given the amplitude required to explain the PTA signal, by extrapolating the CT we would infer $\fp \lesssim 60 \nHz$.
By accounting for the relation between the peak frequency and the duration of the transition $\beta^{-1}$, this would force us to 
$\beta/H_\star \lesssim 1.3 \cdot (150\MeV/T_\star)$.
On top of this upper bound, there is a lower bound imposed by the our assumption about the CT lying around the QCD transition, $T_\star>T_\QCD$. 
Given that $\beta>H_\star$, these equations would imply $T_\star \simeq [150,200] \MeV$.

This back-of-the-envelope estimate is, however, overly restrictive for various reasons.\\ 
{\it i)} First, the parameter space for our assumptions to be consistent becomes larger if one considers 
phase transitions with non-relativistic bubble wall velocities $v_w<1$. Smaller velocities would lead to an additional scaling
of the peak frequency that follows $\fp\propto 0.35/(1+0.07 v_w+0.69 v_w^4)$ or $\fp = 0.536 /v_w$ for the bubble or sound waves contribution respectively.\\
{\it ii)} Secondly, the approximately cubic scaling dictated by causality arguments is expected to break down before the peak frequency, and the SGWB is much less steep between $f_\ct$ and $\fp$.
This alleviates the upper bound from $\Delta \Neff$.
\begin{figure}[h!]\centering
\includegraphics[width=0.5\textwidth]{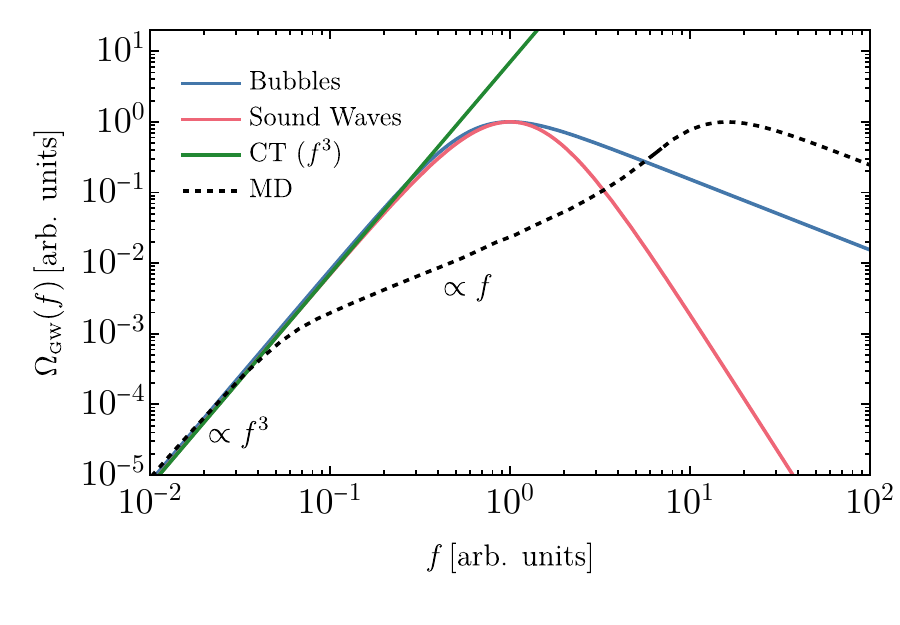}
\vspace{-3em}
\caption{Spectra of SGWB induced by first order PT.
We normalise the frequency in terms of the peak frequency $\fp$ while factorise out the overall spectral amplitude. 
We show both the expected contribution of bubble collisions and sound waves. 
In green we indicate the causality tail obtained assuming perfect radiation domination, while the black dashed line indicates the effect of an intervening matter domination (MD) phase that would shift the peak to larger frequencies.}
\label{fig:fbreaking}
\end{figure}
We show this possibility in Fig.~\ref{fig:fbreaking}, where we adopt the following parametrisation for the contribution of 
bubble collisions \cite{Jinno:2016vai}
\begin{equation}
   \OGW(f) \propto \frac{(a+b)^c}{\left[bx^{-a/c}+ax^{b/c}\right]^c},
\end{equation}	
with $(a,b,c)=(3,1,1.5)$ \cite{NANOGrav:2021flc}
and sound waves \cite{Hindmarsh:2017gnf}
\begin{equation}
   \OGW(f) \propto  x^3\parfrac{7}{4+3x^2}^{7/2}.
\end{equation}
Here, for simplicity, we are neglecting the effects of the QCD crossover on the CT, as they do not affect this discussion.
Due to the breaking of the $f^3$ scaling close to the peak, $\OGW$ is lower than the one obtained extrapolating the CT up to peak frequency of a factor $\mathcal O(6)$. 
\\
{\it iii)} Lastly, an additional model dependence is still possible.
In particular, as also discussed in App.~\ref{app:MD}, 
one may envision an intermediate matter dominated phase that could effectively break the cubic scaling with an intermediate linear growth $\OGW (f) \propto f$.
As a consequence, depending on the duration of the intermediate matter phase, the bound on the SGWB amplitude in the CT could be relaxed, effectively shifting the onset of the cubic drop-off. 

%%%%%%%%%%%%%%%%%%%%%%%%%%%%%%%%%%%%
\section{Effects of intermediate matter domination}
\label{app:MD}
%%%%%%%%%%%%%%%%%%%%%%%%%%%%%%%%%%%%

The interpretation of our results presented in the discussion in the main text are based on Eqs.~\eqref{eq:pfreq} and \eqref{eq:omegapeak}, which assume a standard radiation-dominated (RD) Universe until matter-radiation equality. Changes to the expansion history may have occurred before Big Bang Nucleosynthesis, and would alter the GW relic abundance, as well as the peak frequency and the CT.

The simplest example is arguably that of an early phase of matter domination (MD) when or right after the GW source is active, such as the one that can be due to the slow decay of a massive particle. 
For the purpose of understanding the effects on the GW signal, we parametrize the duration of the MD phase by the ratio of the initial temperature $T_i$ at the onset of MD (if it occurs below $T_\star$, otherwise $T_i=T_\star$) and that at the beginning of the subsequent phase of RD, $r\equiv T_{f}/T_i < 1$. As mentioned in Sec.~\ref{sec:causality tail}, the CT can be dramatically altered in this case, and increases with frequency as $f$.
Notice that the effects of the QCD crossover are in this case washed out, since the SM radiation bath is necessarily a subleading component of the Universe in this scenario.

To model the CT in the presence of a transition from a RD era to a matter dominated (MD) era, we use the following broken power-law parametrization for the characteristic strain:
%\begin{multline}
\begin{equation}
h_{c,\text{MD}}(f)
=A_\ct\left(\frac{f}{f_{\text{yr}}}\right)^{\frac{3-\gamma_{\text{MD}}}{2}} 
%\\ \times 
\left[1+\left(\frac{f}{f_{b,\text{MD}}}\right)^{\frac{1}{\kappa_\text{MD}}}\right]^{\kappa_\text{MD}-\delta_\text{MD}},
\end{equation}
%\end{multline}
with $\gamma_{\text{MD}}=4$ (low frequency, MD) and $\delta_\text{MD}=2$ (high frequency, RD). The bending frequency $f_{b,\text{MD}}$ corresponds to the temperature of the transition from radiation to matter domination.
The choice $\kappa_\text{MD}=0.4$ reproduces the result of~\cite{Hook:2020phx} well enough for our purposes. 

\begin{figure}[h!]
\begin{subfigure}[t]{.62\textwidth}\centering
\includegraphics[width=0.85\textwidth]{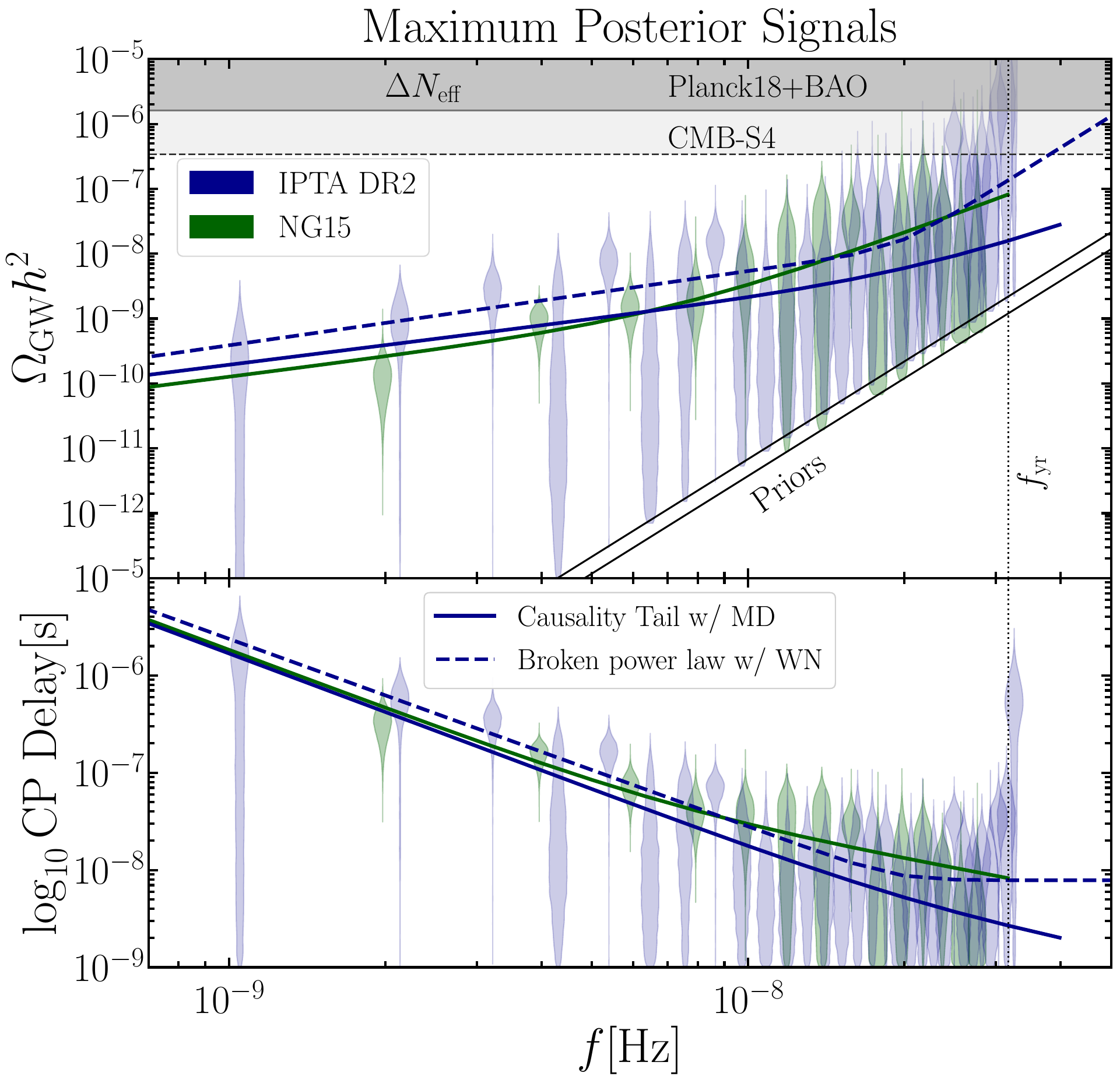}
\caption{Similarly to Fig.~\ref{fig:bestfit}, we show the best fits to IPTA-DR2 data with two different models: in dashed, the broken power-law model with white noise at high frequencies, with values of  bending frequencies determined by our Bayesian analyses; in solid, the CT signals corresponding to intermediate phase of matter domination. For this latter case we also show results obtained using the first 14 bins of the NG15 dataset, including HD correlations. 
The violin-shaped  posteriors are adapted from the results of~\cite{Antoniadis:2022pcn} (IPTA-DR2) and~\cite{NG15-SGWB} (NG15), with lower limits determined by prior choices as in Fig.~\ref{fig:bestfit}.}
\label{fig:bestfit-MD}
\end{subfigure}\hfill%
\begin{subfigure}[t]{.36\textwidth}\centering
\includegraphics[width=\textwidth]{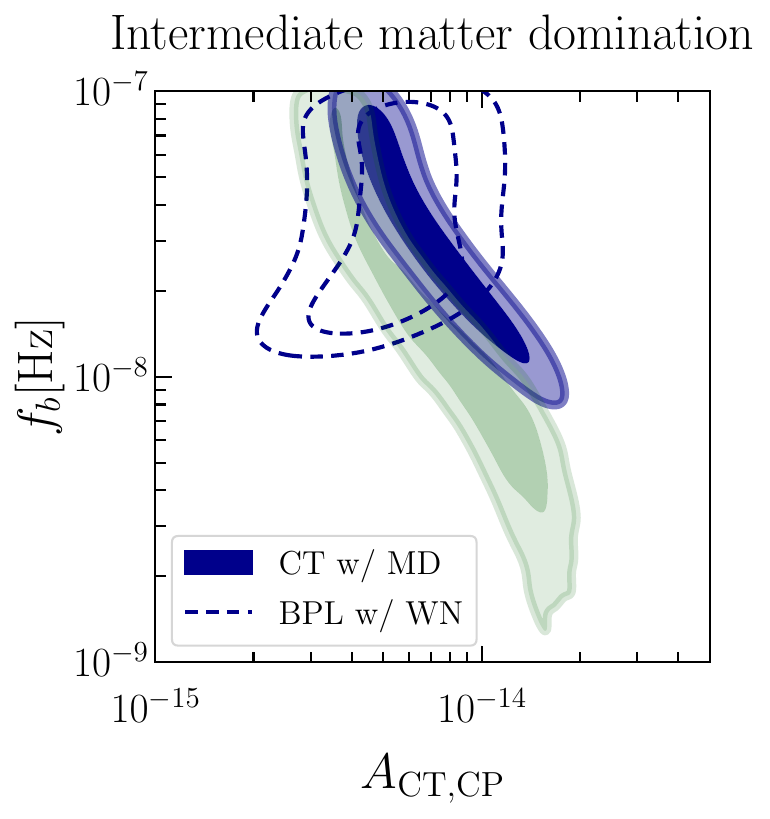}
\caption{2d posteriors of the amplitude of the CT with MD (Broken PL with white noise) and the bending frequency of the signals, for the IPTADR2 dataset (blue). Posteriors for the MD analysis with NG15 are also shown (green).}
\label{fig:MDpost}
\end{subfigure}
\caption{Posteriors and best fits for the scenario of intermediate matter domination.}
\end{figure}

We performed a Bayesian search for such a MD-altered CT in the IPTA-DR2 dataset, additionally including the effects of free-streaming GWs as in \eqref{eq:hCT} of the main text, see also the discussion in the second paragraph of Sec.~\ref{sec:bayes}. To illustrate how this scenario of MD-altered CT signal would compare with data, we show the curves obtained with maximum-posterior values of $f_{b, \text{MD}}$ and $A_{\ct}$ for IPTA-DR2 (this time including all 30 bins, so we set $f_\ct$ larger than the last bin) in Fig.~\ref{fig:bestfit-MD}. We also show the result of a similar analysis in the NG15 dataset, restricting to 14 bins, and including HD correlations (green).
For the bending frequency, the value used in Fig.~\ref{fig:bestfit-MD} for IPTADR2 is 25 nHz, corresponding to $T_i\simeq 800$ MeV, while for NG15 we used $\simeq 7~\text{nHz}$, corresponding to $T_i\simeq 300$ MeV.
For comparison, we also plot the model-agnostic broken power-law with white noise in $A_\cp$ at high frequencies, as in the analyses of the IPTA-DR2 collaborations \cite{Antoniadis:2022pcn} (the curves are obtained by letting the bending frequency, as well as the low-frequency slope and the amplitude free to vary, notice that the difference of our results with respect to those of the collaboration is due to the fact that we are using the maximum posterior values of each parameter, rather than the maximum likelihood signals).

We then compare the models in the IPTA-DR2 dataset with the choices discussed above. The resulting Bayes factor is $\log_{10} \mathcal B_{\text{CT+MD, BPL}}\simeq 1.2$, showing that a CT signal affected by matter domination below the QCD crossover fits the full IPTA-DR2 30 bins dataset better than a broken power-law with white noise at high frencies, although the latter might still be considered as a more motivated model. The 2d posteriors resulting from fitting separately the two models to the data are shown in Fig.~\ref{fig:MDpost}. We find: $\log_{10}A_{\mathrm{CT}}=-14.15^{+0.13}_{-0.21}, \log_{10}f_{b, \mathrm{MD}}=-7.51^{+0.27}_{-0.34}$ and $\log_{10}f_{b, \mathrm{BPL}}=-7.47^{+0.42}_{-0.35}, \gamma_{\text{BPL}}=3.87\pm 0.36$, where the error bars denote the $68\%$ C.L. ranges. For NG15, we find: $\log_{10}A_{\mathrm{CT}}=-14.07^{+0.27}_{-0.38}, \log_{10}f_{b, \mathrm{MD}}=-7.97^{+0.3}_{-0.43}$.

Following the above considerations, 
it is important to notice that additional suppression factors 
in both the peak frequency~\eqref{eq:pfreq}, $\sim r^{1/3}$, and amplitude in \eqref{eq:omegapeak}, $~r^{4/3}$ (if the source is active during MD) arise. Requiring the CT to be in the first bins of the PTA sensitivity band implies that its amplitude is $O(10^{-7}-10^{-8})$ at $f_\yr$. 
In turn, this requires $\alpha\gtrsim (0.1-0.3)r^{-2/3}$. On the other hand, successful BBN imposes $T_f\gtrsim 5\MeV$, i.e.~$r\gtrsim 0.005$ for $f_{\star}\simeq 10^{-7}\Hz$, with the inequality becoming weaker (stronger) if $f_{\star}$ is larger (smaller). 
Therefore the GW source could have overcome the radiation background, if the MD phase lasted until temperatures corresponding to the first frequency bins of PTA datasets.

\begin{figure}[h!]\centering
\includegraphics[width=0.4\textwidth]{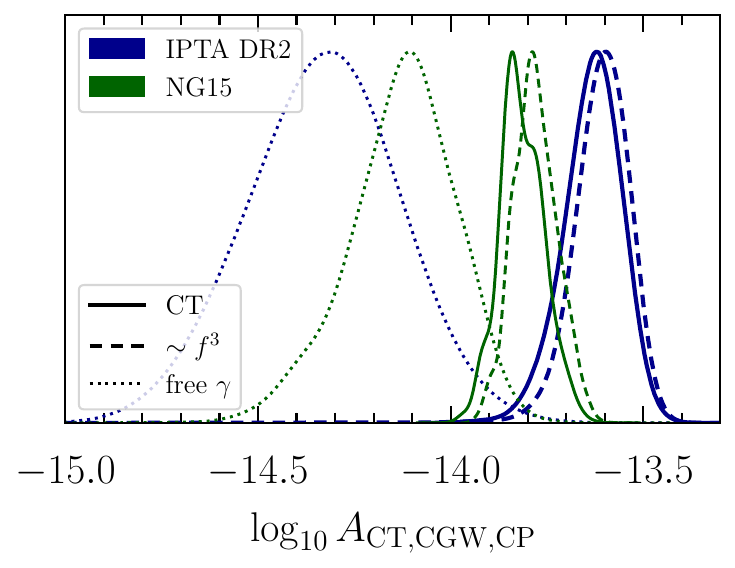}
\caption{1d posteriors of the amplitude of the CT, $f^3$ and free $\gamma$ (CP) signals, obtained using the IPTA-DR2 (blue) or NG15 (green) datasets.}
\label{fig:ACTpost}
\end{figure}

\begin{table*}[t]
\resizebox{\textwidth}{!}{%
\begin{tabular} {lcccccccc}
\toprule
\textit{Parameter} & \multicolumn{3}{c}{\textit{Description}} & \multicolumn{2}{c}{\textit{Prior}} & \multicolumn{3}{c}{\textit{Comments}} \\
\midrule
\multicolumn{9}{c}{\textbf{White Noise}}\\
\midrule
$E_k$& \multicolumn{3}{c}{EFAC per backend/receiver system} & \multicolumn{2}{c}{Uniform $[0,10]$} & \multicolumn{3}{c}{single-pulsar only} \\
$Q_k [s]$& \multicolumn{3}{c}{EQUAD per backend/receiver system} & \multicolumn{2}{c}{log-Uniform $[-8.5,-5]$} & \multicolumn{3}{c}{single-pulsar only} \\
$J_k [s]$& \multicolumn{3}{c}{ECORR per backend/receiver system} & \multicolumn{2}{c}{log-Uniform $[-8.5,-5]$} & \multicolumn{3}{c}{single-pulsar only (NG9)} \\
\midrule
\multicolumn{9}{c}{\textbf{Red Noise}}\\
\midrule
$A_{\text{red}}$& \multicolumn{3}{c}{Red noise power-law amplitude} & \multicolumn{2}{c}{log-Uniform $[-20,-11]$} & \multicolumn{3}{c}{one parameter per pulsar} \\
$\gamma_{\text{red}}$& \multicolumn{3}{c}{Red noise power-law spectral index} & \multicolumn{2}{c}{Uniform $[0,7]$} & \multicolumn{3}{c}{one parameter per pulsar} \\
\midrule
\multicolumn{9}{c}{\textbf{DM Variations Gaussian Process Noise}}\\
\midrule
$A_{\text{DM}}$& \multicolumn{3}{c}{DM noise power-law amplitude} & \multicolumn{2}{c}{log-Uniform $[-20,-11]$} & \multicolumn{3}{c}{one parameter per pulsar} \\
$\gamma_{\text{DM}}$& \multicolumn{3}{c}{DM noise power-law spectral index} & \multicolumn{2}{c}{Uniform $[0,7]$} & \multicolumn{3}{c}{one parameter per pulsar} \\
\midrule
\multicolumn{9}{c}{\textbf{Causality Tail}}\\
\midrule
$A_\ct$& \multicolumn{3}{c}{Strain amplitude at $f=f_\yr$} & \multicolumn{2}{c}{log-Uniform $[-18,$ set by Eq.~\eqref{eq:neffbound}}] & \multicolumn{3}{c}{one parameter per PTA} \\
\midrule
\multicolumn{9}{c}{$\bf{f^3}$}\\
\midrule
$A_{\text{CGW}}$ & \multicolumn{3}{c}{Strain amplitude at $f=f_\yr$} & \multicolumn{2}{c}{log-Uniform $[-18,$ set by Eq.~\eqref{eq:neffbound}}] & \multicolumn{3}{c}{one parameter per PTA} \\
\midrule
\multicolumn{9}{c}{\textbf{Fixed power-law}}\\
\midrule
$A_\text{SMBHB}$& \multicolumn{3}{c}{Strain amplitude at $f=f_{\text{yr}}$} & \multicolumn{2}{c}{log-Uniform $[-18, -11]$} & \multicolumn{3}{c}{one parameter per PTA}\\
\midrule
\multicolumn{9}{c}{\textbf{power-law}}\\
\midrule
$A_{\text{GWB}}$& \multicolumn{3}{c}{Strain amplitude at $f=f_{\text{yr}}$} & \multicolumn{2}{c}{log-Uniform $[-18, -11]$} & \multicolumn{3}{c}{one parameter per PTA}\\
$\gamma_{\text{GWB}}$& \multicolumn{3}{c}{GWB power-law spectral index} & \multicolumn{2}{c}{Uniform $[0,7]$} & \multicolumn{3}{c}{one parameter per PTA}\\
\midrule
\multicolumn{9}{c}{\textbf{Matter domination CT}}\\
\midrule
$A_\text{CT}$& \multicolumn{3}{c}{Strain amplitude at $f=f_{\text{yr}}$} & \multicolumn{2}{c}{log-Uniform $[-18,-11]$} & \multicolumn{3}{c}{one parameter per PTA}\\
$\gamma_{\text{MD}}$& \multicolumn{3}{c}{CT low frequency power-law spectral index} & \multicolumn{2}{c}{Fixed to $4$} & \multicolumn{3}{c}{one parameter per PTA}\\
$\delta_{\text{MD}}$& \multicolumn{3}{c}{CT high-frequency power-law spectral index} & \multicolumn{2}{c}{Fixed to $2$} & \multicolumn{3}{c}{one parameter per PTA}\\
$f_{b, \text{MD}}$& \multicolumn{3}{c}{CT bending frequency (MD to RD transition)} & \multicolumn{2}{c}{log-Uniform $[-10, -7]$} & \multicolumn{3}{c}{one parameter for PTA}\\
$\kappa_{\text{MD}}$& \multicolumn{3}{c}{Width of matter to radiation transition} & \multicolumn{2}{c}{Fixed to $0.4$} & \multicolumn{3}{c}{one parameter per PTA}\\
\midrule
\multicolumn{9}{c}{\textbf{High frequency noise Broken power-law}}\\
\midrule
$A_{\text{GWB}}$& \multicolumn{3}{c}{Strain amplitude at $f=f_\yr$} & \multicolumn{2}{c}{log-Uniform $[-18, -11]$} & \multicolumn{3}{c}{one parameter per PTA}\\
$\gamma_{\text{GWB}}$& \multicolumn{3}{c}{GWB low frequency power-law spectral index} & \multicolumn{2}{c}{Uniform $[0,7]$} & \multicolumn{3}{c}{one parameter per PTA}\\
$\delta_{\text{GWB}}$& \multicolumn{3}{c}{CT high-frequency power-law spectral index} & \multicolumn{2}{c}{Fixed to $0$} & \multicolumn{3}{c}{one parameter per PTA}\\
$f_{\text{b}}$& \multicolumn{3}{c}{GWB bending frequency (signal to noise transition)} & \multicolumn{2}{c}{log-Uniform $[-10, -7]$} & \multicolumn{3}{c}{one parameter per PTA}\\
$\kappa_{\text{GWB}}$& \multicolumn{3}{c}{Width of signal to noise transition} & \multicolumn{2}{c}{Fixed to: $0.1$} & \multicolumn{3}{c}{one parameter per PTA}\\
\bottomrule
\end{tabular} %
}
  \caption{List of noise and GWB parameters used in our analyses, together with their prior ranges.}
  \label{table:priors}
\end{table*}
%

%%%%%%%%%%%%%%%%%%%%%%%%%%%%%%%%%%%%%%%%%%%%%%%%%%%%
\section{Details of the numerical analysis and additional posteriors}
\label{app:priors}
%%%%%%%%%%%%%%%%%%%%%%%%%%%%%%%%%%%%%%%%%%%%%%%%%%%%

Here we report details of our numerical analysis.  The analysis to obtain Bayes factors has been performed using using the codes {\tt enterprise}~\cite{enterprise} and {\tt en\-ter\-pri\-se\_extensions}~\cite{enterprise_ext}, in which we implemented the CT signal reported in Fig.~\ref{fig:QCD-PT}, and use {\tt PTMCMC}~\cite{justin_ellis_2017_1037579} to obtain MonteCarlo samples. We derive posterior distributions using {\tt GetDist}~\cite{Lewis:2019xzd}. We include noise parameters (white, red and dispersion measure) according to the strategy of the IPTA DR2~\cite{Antoniadis:2022pcn}. For more details, see also the Appendix of~\cite{Ferreira:2022zzo}. For the NG15 dataset, we have used {\tt PTArcade}~\cite{Mitridate:2023oar}.

Our prior choices for noise (DM only applies to IPTA DR2) and GW parameters are reported in Table~\ref{table:priors}. In order to produce the results reported in the main text, we have restricted our analyses to the first 13 and 14 frequency bins of the IPTA-DR2 and NG15 datasets respectively, as in~\cite{Antoniadis:2022pcn, NG15-SGWB}. We obtained more than $10^6$ (typically $5\cdot 10^6$) samples per each analysis and model in this work, and discarded the first $25\%$ of each chain. With these choices, we reproduce the posteriors for CP parameters presented in~\cite{Antoniadis:2022pcn} with excellent agreement.

The upper prior boundary on the amplitude of cosmological signals is set by the $\Delta \Neff$ bound reported in Eq.~\eqref{eq:neffbound}, setting $f_\ct = f_\yr$ for the analyses reported in the main text to set ideas. 

For the NG15 dataset, the Bayes factor with respect to SMBHBs model has been computed using the prior on the astrophysical signal provided in {\tt PTArcade}, which is based on population synthesis assuming circular orbits and energy loss due only to GWs.

We employed all 30 frequency bins of the IPTA-DR2 datasets for the intermediate MD analysis presented in App.~\ref{app:MD}.

The maximum-posterior curves for the NG15 dataset in Fig.~\ref{fig:bestfit} and Figs.~\ref{fig:bestfit-MD} and \ref{fig:MDpost} have been obtained using the mode {\tt ceffyl} in {\tt PTArcade}. We fixed $\log_{10} f_\ct=-7.1 (-7.5)$ for the IPTADR2 (NG15) intermediate MD analysis. The faster fitting procedure allowed us to include HD correlations.

We show the 1d posteriors for the amplitude of the GW signals considered in Sec.~\ref{sec:bayes} of the main text in Fig.~\ref{fig:ACTpost}. We find: for IPTA-DR2, $\log_{10}A_{\mathrm{CT}}=-13.63^{+0.08}_{-0.07}, \log_{10}A_{f^3}=-13.61^{+0.08}_{-0.06}$ and $\log_{10}A_{\mathrm{CP}}=-14.33\pm 0.2, \gamma = 3.94\pm 0.44$ for the CP signal with free $\gamma$, where the error bars denote the $68\%$ C.L. ranges. For NG15: $\log_{10}A_{\mathrm{CT}}=-13.82\pm{0.06}, \log_{10}A_{f^3}=-13.781^{+0.05}_{-0.07}$ and $\log_{10}A_{\mathrm{CP}}=-14.13^{+0.16}_{-0.12}, \gamma = 3.09^{+0.33}_{-0.38}$.

\end{document}